\documentclass[12pt,sec]{article}
\usepackage{epsfig}
\usepackage{axodraw}
\begin{document}

\begin{center}
{\large S-D mixing and $\psi(3770)$ production in $e^+e^-$ annihilation and B decay and its radiative transitions}\\[0.8cm]

{ Kui-Yong Liu}

{\footnotesize Department of Physics, Peking University,
 Beijing 100871, People's Republic of China
 and Department of Physics, Liaoning University, Shenyang 110036, People's Republic of China}\\[0.5cm]

{ Kuang-Ta Chao}

{\footnotesize China Center of Advanced Science and Technology
(World Laboratory), Beijing 100080, People's Republic of China and
Department of Physics, Peking University,
 Beijing 100871, People's Republic of China}

\end{center}

\begin{abstract}

The large decay rate observed by Belle for
$B^+\rightarrow\psi(3770)K^+$, which is comparable to
$B^+\rightarrow\psi(3686)K^+$, might indicate either an
unexpectedly large S-D mixing angle $|\theta|\approx 40^o$ or the
leading role of the color-octet mechanism in D-wave charmonium
production in $B$ decay. By calculating the production rate of
$\psi(3770)$ in the continuum $e^+e^-$ annihilation at
$\sqrt{s}=10.6$ GeV with these two possible approaches (i.e. the
large S-D mixing and the color-octet mechanism), we show that the
measurement for this process at Belle and BaBar may provide a
clear cut clarification for the two approaches. In addition, the
radiative E1 transition ratio $\Gamma(\psi(3770)\rightarrow
\gamma\chi_{c2})/\Gamma(\psi(3770)\rightarrow \gamma\chi_{c1})$
may dramatically change from $\sim$ 0.04 (for $\theta\approx 0^o$)
to $\sim$ 200 (for $\theta\approx -40^o$) due to the large S-D
interference effect, thus the E1 transition measurement of
$\psi(3770)$ at BES and CLEO-c will also be very useful in
clarifying this issue.

PACS number(s): 13.66.Bc, 14.40.Gx, 12.38.Bx

\end{abstract}

\section{Introduction}

The S-D mixing for $\psi'\equiv\psi(3686)$ and
$\psi''\equiv\psi(3770)$ is of great interest in charmonium
physics. If we neglect the charmed meson pair component which is
due to coupling to decay channels, $\psi'$ and $\psi''$ may be
approximately expressed as
\begin{equation}
\mid \psi' \rangle = cos\theta \mid 2^3S_1\rangle + sin\theta
\mid 1^3D_1\rangle, ~~~~~    \\
\mid \psi'' \rangle = cos\theta \mid 1^3D_1\rangle + sin\theta
\mid 2^3S_1\rangle.
\end{equation}
A rough estimate of the S-D mixing angle may be obtained by using
the ratio of the observed leptonic decay widths \cite{pdg} of
$\psi'$ and $\psi''$ and neglecting the D-wave component
contribution to these leptonic decay widths ${\rm
tan}^2\theta=\frac{\Gamma(\psi''\rightarrow
l^+l^-)}{\Gamma(\psi'\rightarrow l^+l^-)}\approx 0.12$, which
results in $\theta\approx \pm 19^o$. However, if the D-wave
contribution to leptonic decay widths is further included,
potential model calculations e.g. in \cite{kuang,ding,rosner} give
two solutions: $\theta\approx -10^o$ to $-13^o$ or $\theta\approx
+30^o$ to $+26^o$ ($\theta\approx -10^o$ and $+30^o$ in
\cite{kuang},  $\theta\approx -13^o$ and $+26^o$ in \cite{ding},
and $\theta\approx -12^o$ and $+27^o$ in \cite{rosner}). The small
mixing solution (i.e. $\theta\approx -10^o$) is compatible with
the results obtained in models with coupled decay channels
\cite{eichten, heikkila}. Moreover, $\theta\approx -10^o$ is
favored by the $\psi'\rightarrow \gamma\chi_{cJ}$ data whereas
$\theta\approx +30^o$ would lead to $\Gamma(\psi'\rightarrow
\gamma\chi_{c0})\approx $135 KeV due to a large positive S-D
interference \cite{ding}, which is higher than its observed value
by a factor of 6 and is therefore disfavored. The S-D mixing may
have many interesting phenomenological consequences. It might
slightly \cite{kuang} or substantially \cite{kwong,moxhay} affect
the $\psi(3770)\rightarrow J/\psi\pi\pi$ decay , which has been
recently observed by BES \cite{bes}. It would also have effects on
the $h_c$ search via the $\psi'\rightarrow h_c\pi^0$ decay
\cite{kuang1}. The S-D mixing could even provide an explanation
for the notorious $\rho\pi$ puzzle that the suppression of
$\psi'\rightarrow\rho\pi$ is due to a destructive interference
between the S and D wave states \cite{rosner}, and it might also
be useful in explaining \cite{wangping} the recent observed
enhancement of $\psi'\rightarrow K_LK_S$ by BES \cite{bes1}.
Moreover, the ratio of $\frac{\Gamma(\psi(3770)\rightarrow
\gamma\chi_{c2})}{\Gamma(\psi(3770)\rightarrow \gamma\chi_{c1})}$
may sensitively change from 0.04 (for $\theta\approx 0^o$) to 0.22
(for $\theta\approx -10^o$) and to 0.06 (for $\theta\approx
+30^o$) \cite{ding}). The experimental examination of these
radiative transitions for $\psi(3770)$ at BES and CLEO-c in the
near future will be an interesting test for the S-D mixing.

Recently Belle Collaboration \cite{belled} has observed
$\psi(3770)$ for the first time in the B meson decay
$B^+\rightarrow\psi(3770)K^+$ with a branching ratio of $(0.48\pm
0.11\pm 0.07)\times 10^{-3}$, which is comparable to ${\cal
B}(B^+\rightarrow\psi'(3686)K^+)=(0.66\pm 0.06)\times 10^{-3}$
\cite{pdg}. This is quite surprising, since conventionally
$\psi(3770)$ and $\psi(3686)$ are regarded as mainly the $1^3D_1$
and $2^3S_1$ color-singlet $c\bar c$ states respectively, and the
coupling of $1^3D_1$ to the $c \bar c$ vector current in the weak
decay effective hamiltonian is much weaker than that of $2^3S_1$.
One possible explanation is that the S-D mixing for $\psi(3770)$
and $\psi(3686)$ is very large, much larger than previously
expected . If in the $B$ meson decay we neglect the $1^3D_1$
contribution, which is expected to be much smaller than the
$2^3S_1$ contribution, we would get the S-D mixing angle
$|\theta|\approx 40.4^o$ from the observed decay rate ratio
\begin{equation}
\frac{{\cal B}(B^+\rightarrow\psi(3770)K^+)}{{\cal
B}(B^+\rightarrow\psi(3686)K^+)} =\tan^2\theta\approx 0.73.
\end{equation}
Although this large mixing angle seems to be not compatible with
all our previous knowledge about the S-D mixing,  it is still
worthwhile to test it with new experiments. Another possible
explanation is that the color-octet mechanism in nonrelativistic
QCD (NRQCD) may play the leading role in the D-wave charmonium
production in B meson decays, and it was predicted \cite{yuan}
that for a pure D-wave state the branching ratio ${\cal B}(
B\rightarrow\psi(3770)X)\approx 0.28\% $, which is comparable to
${\cal B}( B\rightarrow\psi(3686)X)= (0.35\pm 0.05)\%$ \cite{pdg}
(see also \cite{ko1} for similar discussions). In a series of
papers \cite{qiao1,qiao2,yuan,yuan1} it was pointed out that based
on the Fock state expansion and velocity scaling rules of NRQCD
\cite{bbl} the production rates of D-wave charmonium states
including the $^3D_1~\psi(3770)$, which would predominantly decay
to the $D\bar D$ meson pair, and an expected narrow $^3D_2$ state,
which could have some decay fraction to $J/\psi\pi^+\pi^-$, would
be comparable to that of $J/\psi$ and $\psi(2S)$ in the $Z^0$
decay \cite{qiao1}, the $p\bar p$ collision at the Tevatron
\cite{qiao2}, the B meson decay \cite{yuan}, and the fixed target
experiments \cite{yuan1}, whereas in the conventional
color-singlet model the D-wave production rates, which are
proportional to the squared second derivative of the $c\bar c$
wave function at the origin, should be greatly suppressed. Despite
of certain uncertainties associated with the values of color-octet
matrix elements, the color-octet contributions are expected to be
dominant, which are larger than the color-singlet contributions by
more than one order of magnitude in those processes. For instance,
in the large $p_T$ charmonium production at the Tevatron, the
$^3D_J$ states could have large rates, comparable to $\psi(2S)$,
because in both cases the dominant production mechanism is
expected to be gluon fragmentation into the color-octet $^3S_1$
$c\bar c$ intermediate state which then evolves respectively into
the physical color-singlet $^3D_J$ and $\psi(2S)$ states by
emitting two soft gluons via double E1 transitions with the same
order transition probabilities \cite{qiao2}. Similarly, the
$\psi(3770)$ production in the B meson decay could provide another
interesting test for the color-octet mechanism in NRQCD.

In order to distinguish between the large S-D mixing and the NRQCD
color-octet mechanism in the $\psi(3770)$ production in the B
meson decay, we suggest measuring the $\psi(3770)$ production in
continuum $e^+e^-$ annihilation at the Belle and BaBar energy
$\sqrt{s}=10.6$ GeV, and we will give calculations of the
production cross sections in the following sections. We will show
that the $\psi(3770)$ production in  $e^+e^-$ annihilation via
double $c\bar c$ does not receive large contributions from the
color-octet mechanism but is very sensitive to the S-D mixing.

$J/\psi$ inclusive production in $e^+e^-$ annihilation has been
investigated within the color-singlet model \cite{cm3} and the
color-octet model \cite{om1,om2,ko}. BaBar \cite{babar} and Belle
\cite{belle} have measured $J/\psi$ production rate in continuum
$e^+e^-$ annihilations at $\sqrt{s}=10.6 GeV$, which is found to
be much larger than the color-singlet prediction. More
interestingly, Belle further finds $J/\psi$ production to be
dominated by the double $c\bar c$ production \cite{exdou}. The
measured exclusive cross section for $e^+ + e^-\rightarrow
J/\psi+\eta_c$ is an order of magnitude larger than the
theoretical values\cite{double}, and the measured inclusive cross
section for $e^+ + e^-\rightarrow J/\psi+c+\bar c$  is more than
five times larger than NRQCD predictions. Therefore the inclusive
charmonium production via double $c\bar c$ is particularly
interesting and worth investigating. As in the case of inclusive
production of $J/\psi$, $\eta_{c}$, and $\chi_{cJ}$(J=0, 1,
2)\cite{liu1}, here we will also concentrate on the double $c\bar
c$ production for the D-wave charmonium states
$\delta_{J}(J=1,2,3)$.

\section{ Color-singlet contribution
to $\delta_1$ ($^3D_1$) charmonium production via double
$c\bar{c}$ in $e^+e^-$ annihilation}

In NRQCD the Fock state expansion for the D-wave (without S-D
mixing) charmonium $\delta_J (J=1,2,3)$ is
\begin{eqnarray}
\label{expandfock} \nonumber |\delta_J\rangle
&&=O(1)|c\bar{c}({}^{3}D_{J},\underline{1}) \rangle\\
\nonumber
&& +O(v)|c\bar{c}({}^{3}P_{J'},\underline{8})g\rangle\\
&& +O(v^2)|c\bar{c}({}^{3}S_1,\underline{8})~ gg\rangle+\cdots.
\end{eqnarray}
Following the NRQCD factorization formalism, the scattering
amplitude of the process $e^-(p_1)+e^+(p_2)\rightarrow \gamma^*
\rightarrow
c\bar{c}(^{2S+1}L_J^{(1,8a)})(p)+c(p_c)+\bar{c}(p_{\bar{c}})$ in
Fig.~\ref{feynman} is given by
\begin{eqnarray}
%\hspace{-1.0cm}\hspace{1.0cm}
\label{amp2}   &&\hspace{-1cm}{\cal
A}(e^-(p_1)+e^+(p_2)\rightarrow
c\bar{c}(^{2S+1}L_{J}^{(1,8a)})(p)+c(p_c)+\bar{c}(p_{\bar{c}}))=\sqrt{C_{L}}
\sum\limits_{L_{z} S_{z} }\sum\limits_{s_1s_2 }\sum\limits_{jk}
\nonumber
\\ & \times&\langle s_1;s_2\mid
S S_{z}\rangle \langle L L_{z};S S_{z}\mid J J_{z}\rangle\langle
3j;\bar{3}k\mid 1,8a\rangle\nonumber\\
&\times&\left\{
\begin{array}{ll}
{\cal A}(e^-(p_1)+e^+(p_2)\rightarrow
 c_j(\frac{p}{2};s_1)+\bar{c}_k(\frac{p}{2};s_2)+
 c_l(\frac{p_c}{2};s_3)+\bar{c}_i(\frac{p_{\bar{c}}}{2};s_4))&(L=S),\nonumber\\
\epsilon^*_{\alpha}(L_Z) {\cal
A}^\alpha(e^-(p_1)+e^+(p_2)\rightarrow
 c_j(\frac{p}{2};s_1)+\bar{c}_k(\frac{p}{2};s_2)+
 c_l(\frac{p_c}{2};s_3)+\bar{c}_i(\frac{p_{\bar{c}}}{2};s_4))
&(L=P),\nonumber\\
\frac{1}{2}\epsilon^*_{\alpha \beta}(L_Z) {\cal A}^{\alpha
\beta}(e^-(p_1)+e^+(p_2)\rightarrow
 c_j(\frac{p}{2};s_1)+\bar{c}_k(\frac{p}{2};s_2)+
 c_l(\frac{p_c}{2};s_3)+\bar{c}_i(\frac{p_{\bar{c}}}{2};s_4))
&(L=D).
\end{array}
\right.\nonumber\\
\end{eqnarray}
where $c\bar{c}(^{2S+1}L_{J}^{(1,8a)})$ is the $c\bar{c}$ pair
produced at short distances, which subsequently evolve into a
specific charmonium state at long distances, ${\cal A}^\alpha$ and
${\cal A}^{\alpha \beta}$ are the derivatives of the amplitude
with respect to the relative momentum between the quark and
anti-quark in the bound state. For the case of color-singlet
state, the coefficient $C_{L}$ can be related to the origin of the
radial wave function (or its derivatives) of the bound state as
\begin{equation}
C_{D}=\frac{15}{8\pi}\mid R_{D}''(0) \mid^{2}.
\end{equation}
The spin projection operators and their derivatives with respect
to the relative momentum are
\begin{equation}
P_{1S_Z}(p,0)=\frac{1}{2\sqrt{2}}\not{\epsilon}(S_z)(\not{p}+2m_{c}),
\end{equation}
\begin{equation}
P_{1S_z}^{\alpha}(p,0)=\frac{1}{4\sqrt{2}m_c}
[\gamma^{\alpha}\not{\epsilon}^*(S_z)(\not{p}+2m_c)-
(\not{p}-2m_c)\not{\epsilon}(S_z)\gamma^{\alpha}].
\end{equation}
\begin{equation}
P_{1S_z}^{\alpha\beta}(p,0)=\frac{1}{2\sqrt{2}m_c}
[\gamma^{\alpha}\not{\epsilon}^*(S_z)\gamma^{\beta}+
\gamma^{\beta}\not{\epsilon}(S_z)\gamma^{\alpha}].
\end{equation}

The calculation of cross sections for $e^{-}+e^{+}\rightarrow
\gamma^*\rightarrow$ charmonium $ + c\bar{c}$ is straightforward.
As in Ref.~\cite{cm3} we write the differential cross section as
\begin{equation}
\label{cross} \frac{d\sigma(e^{+}+e^{-}\rightarrow \gamma^{*}
\rightarrow {\rm charmonium} +
c\bar{c})}{dz}=\frac{4C_{D}\alpha^{2}\alpha_{s}^{2}}{81m_{c}}(S(z)+\frac{\alpha(z)}{3}).
\end{equation}
where $z=2E_{\psi}/\sqrt{s}$. The expressions of S(z) and
$\alpha(z)$ for $\delta_J$ are lengthy and will be given in the
appendix for $J=1,2$. With Eq.~(\ref{cross}) we can evaluate the
inclusive cross sections for $\delta_J$. The input parameters used
in the numerical calculations are
\begin{equation}
m_c=1.5GeV, ~~\alpha_s(2m_c)=0.26,~~\alpha=1/137,~~\mid R_D''(0)
\mid^2=0.015GeV^7\cite{wf}.
\end{equation}
and the obtained cross section for the $\delta_1$ at
$\sqrt{s}=10.6$ GeV is
\begin{equation}
\label{dt1}
\sigma(e^{+}+e^{-}\rightarrow\gamma^{*}\rightarrow\delta_{1}+c\bar{c})=2.5~
{\rm fb}.
\end{equation}
Here we also give the calculated cross section for the $\delta_2$
at $\sqrt{s}=10.6$ GeV
\begin{equation}
\label{dt2}
\sigma(e^{+}+e^{-}\rightarrow\gamma^{*}\rightarrow\delta_{2}+c\bar{c})=2.4~
{\rm fb}.
\end{equation}
We also find that as in the case of other charmonium states
\cite{liu1} the calculated cross sections for $\delta_1$ in
eq.~(\ref{dt1}) and $\delta_2$ in eq.~(\ref{dt2}) are
substantially smaller than those obtained in the fragmentation
approximation at $\sqrt{s}=10.6$ GeV which would cause a
enhancement factor of 1.5 and 2.3 respectively.

Comparing eq.(\ref{dt1}) with
$\sigma(e^{+}+e^{-}\rightarrow\gamma^{*}\rightarrow
J/\psi+c\bar{c})=148~fb$ calculated for the $J/\psi$ in
\cite{liu1}, we see that the inclusive double $c\bar c$ cross
section for the D-wave $1^{--}$ state is smaller than that for the
S-wave $1^{--}$ states by a factor of 60. This illustrates the
expectation that within the color-singlet model the suppression of
D-wave state production relative to the S-wave state production is
usually about two orders of magnitude.

\section{Color-octet contribution to $\delta_1$
production via double $c\bar{c}$ in $e^+e^-$ annihilation}

\begin{figure}
\begin{center}
%\vspace{-2cm}
\includegraphics[width=14cm,height=18cm]{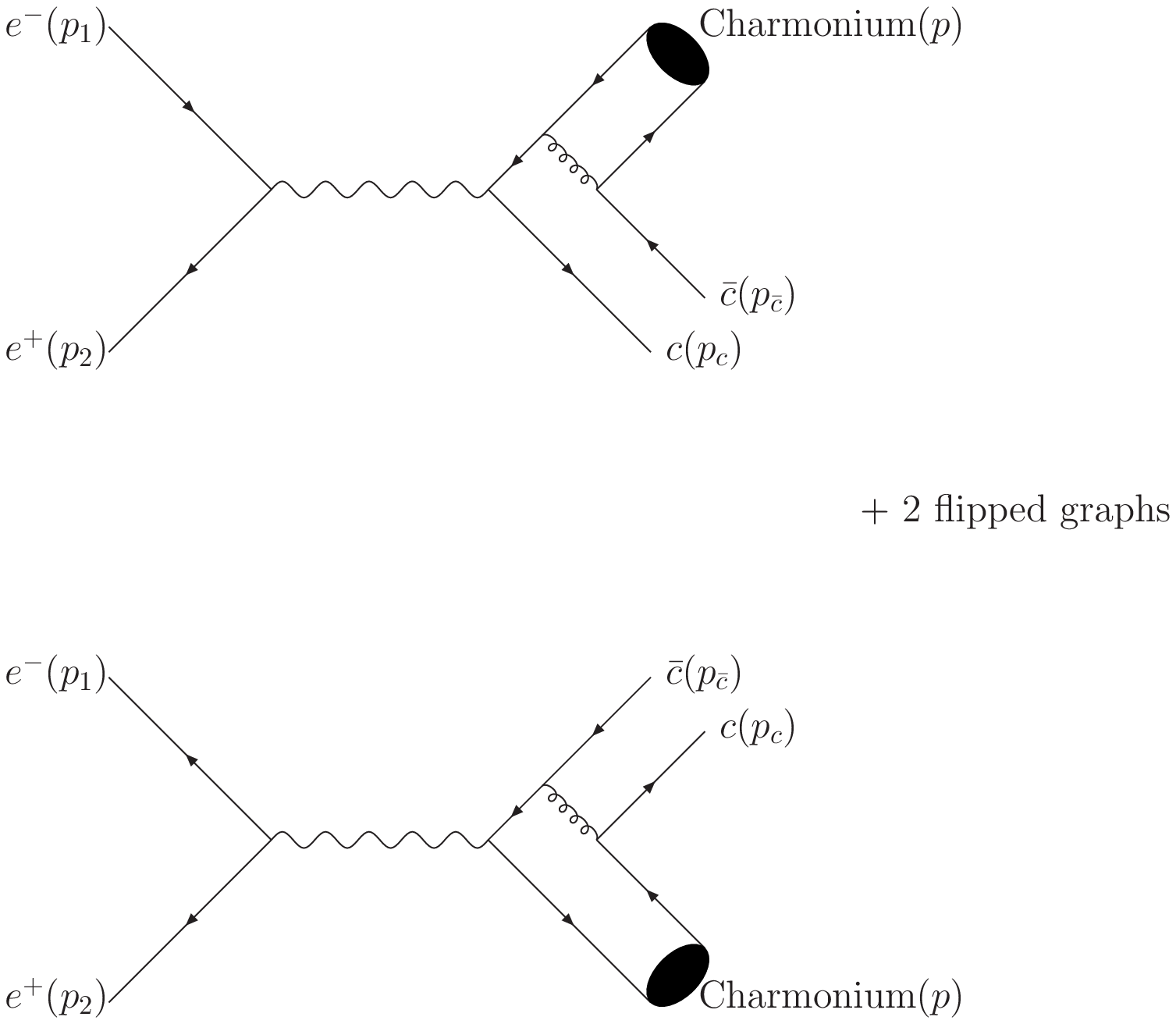}
%\vspace{-4cm}
\end{center}
\caption{ Feynman diagrams for $e^+ +
e^-\rightarrow\gamma^*\rightarrow$ charmonium + $c\bar{c}$.}
\label{feynman}
\end{figure}

\begin{figure}
\begin{center}
\vspace{-3.0cm}
\includegraphics[width=14cm,height=18cm]{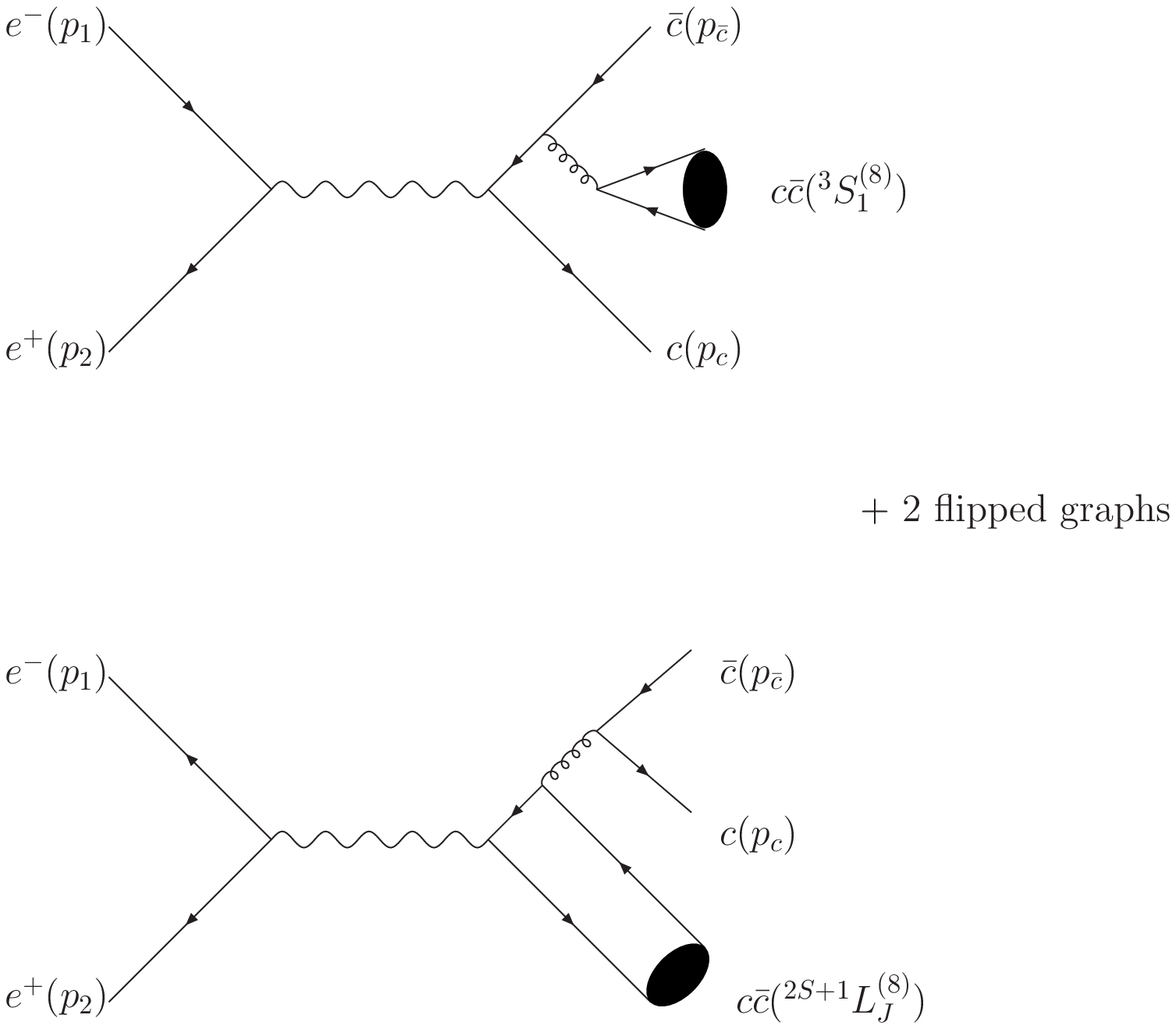}
\vspace{-4cm}
\end{center}
\caption{ Feynman diagrams for $e^+ +
e^-\rightarrow\gamma^*\rightarrow$ $c\bar{c}(^{2S+1}L_J^{(8)})$ +
$c\bar{c}$.} \label{fey2}
\end{figure}

The color-octet contributions to $\delta_1$ production via double
$c\bar{c}$ in $e^+e^-$ annihilation come from the Feynman diagrams
in Fig~\ref{feynman} and Fig~\ref{fey2}. According to the NRQCD
factorization and velocity scaling rules\cite{bbl}, the
contributions of the second term and third term in the Fock state
expansion of $\delta_1$ in eq.~(\ref{expandfock}) are of the same
order in the quark relative velocity $v$ as the corresponding
terms in the Fock state expansion of $J/\psi$ or $\psi(2S)$. As a
rough estimate for the nonperturbative matrix elements we may
choose

\begin{equation}
<{\cal O}^{\delta_1}_8(^{3}S_1)>\approx<{\cal
O}^{J/\psi}_8(^{3}S_1)>=1.06\times 10^{-2}~{\rm GeV}^3,
\end{equation}
\begin{equation}
<{\cal O}^{\delta_1}_8(^{3}P_0)>/m_c^2\approx<{\cal
 O}^{J/\psi}_8(^{3}P_0)>/m_c^2=1.1\times 10^{-2}~{\rm GeV}^3,
\end{equation}
\begin{equation}
<{\cal O}^{\delta_1}_8(^{1}S_0)>\ll<{\cal
O}^{J/\psi}_8(^{1}S_0)>=3.3\times 10^{-2}~{\rm GeV}^3,
\end{equation}
\begin{equation}
<{\cal O}^{H}_8(^{3}P_J)>=(2J+1)<{\cal O}^{H}_8(^{3}P_0)>.
\end{equation}
Here the color-octet matrix elements for $J/\psi$ were extracted
from the $J/\psi$ data at the Tevatron (see Ref.~\cite{omes,omes1}
for detailed discussions). There are large uncertainties with
$<{\cal O}^{J/\psi}_8(^{3}P_0)>/m_c^2 $ and $<{\cal
O}^{J/\psi}_8(^{1}S_0)>/3$ and their combinations, and here we
have assumed that they are equal and take the largest fitted
values from \cite{omes} to avoid underestimates of the color-octet
contributions. (note that these two matrix elements may be
overestimated \cite{omes1}.)

In Fig.~\ref{feynman}, the color-octet contribution can be
obtained from the corresponding color-singlet contribution divided
by a factor of $\frac{32<{\cal O}^{H}_1(^{2S+1}L_{J})>}{3<{\cal
O}^{H}_8(^{2S+1}L_J)>}$. With the matrix elements for $<{\cal
O}^{\delta_1}_8(^{2S+1}L_J)>$ chosen above, we find the
contributions to $\delta_1$ production cross section from the
color-octet $^3S_1, ^3P_0, ^3P_1, ^3P_2$ states to be 0.12fb,
1.1fb, 0.29fb, 0.14fb respectively. The total color-octet
contribution to the cross section from Fig.~\ref{feynman} is
1.65fb.

In Fig.~\ref{fey2} the color-octet contributions come from four
different (the upper two and the lower two) diagrams. The upper
diagrams only contribute via the color-octet $^3S_1$ state, and
the differential cross section reads
\begin{equation}
\frac{d\sigma_{octet}}{dz}=\frac{16\alpha^2\alpha_s^2<{\cal
O}^{\delta_1}_8(^{3}S_1)>}{27m_c}\mid \bar{M} \mid^2,
\end{equation}
where $\mid \bar{M} \mid^2$ takes the form
\begin{eqnarray}
\mid \bar{M} \mid^2&=& \frac{\pi}{12 \delta^2 s^2 z
(z-2)^2}\{-4z\sqrt{\frac{(1-z)(z^2-\delta^2)}{4+\delta^2-4z}}\nonumber\\
&&[3\delta^4-12\delta^2(z-2)+16(10+z(z-10))]+
\nonumber\\
&&(z-2)^2[3\delta^4-8\delta^2(3z-4)+32(2+z(z-2))]\nonumber\\
&&{\rm ln}[\frac{z \sqrt{4 + \delta^2 - 4
z}+2\sqrt{(1-z)(z^2-\delta^2)}}{z \sqrt{4 + \delta^2 - 4
z}-2\sqrt{(1-z)(z^2-\delta^2)}}]\}.
\end{eqnarray}
The numerical result for the cross section is
\begin{equation}
\sigma_{octet}(e^+e^-\rightarrow \delta_1 c\bar{c})=4.5~{\rm fb}.
\end{equation}

The lower diagrams in Fig.~\ref{fey2} make contributions to
$\delta_1$ production via the color-octet $^3P_J (J=0,1,2)$ and
$^1S_0$ intermediate states. We find the following results.
\begin{equation}
\frac{d\sigma_{octet}}{dz}=\frac{32\alpha^2\alpha_s^2<{\cal
O}^{\delta_1}_8(^{2S+1}L_J)>}{27m_c}\sqrt{\frac{(z^2-\delta^2)(1-z)}{4+\delta^2-4z}}\mid
\bar{M}(^{2S+1}L_J) \mid^2,
\end{equation}

\begin{eqnarray}
\mid \bar{M}(^3P_0) \mid ^2=\frac{(16 \pi (8 + 3 \delta^2 - 8 z)
(\delta^6 + 8 \delta^2 (-4 + z) z^2 + 8 z^4 +
        2 \delta^4 (18 + (-10 + z) z)))}{(27 \delta^2 s^3 (4 + \delta^2 - 4 z)^2 (-2 + z)^4)}
\end{eqnarray}

\begin{eqnarray}
\mid \bar{M}(^3P_1) \mid ^2&=&\frac{1}{(9 \delta^2 s^3 (4 +
\delta^2 - 4 z)^2 (-2 + z)^4)}\nonumber\\
&\times&(16 \pi (8 + 3 \delta^2 - 8 z) (\delta^6 + 16 (-2 + z)^2
z^2 + 2 \delta^4 (18 + (-12 + z) z)\nonumber\\
&-& 4 \delta^2 (-2 + z) (16 + z (-20 + 3 z))))
\end{eqnarray}

\begin{eqnarray}
\mid \bar{M}(^3P_2) \mid ^2=\frac{(32 \pi (8 + 3 \delta^2 - 8 z)
(\delta^6 + 2 \delta^4 (-3 + z) (-2 + z) + 4 z^4 -
        2 \delta^2 z^2 (2 + z)))}{(9 \delta^2 s^3 (4 + \delta^2 - 4 z)^2 (-2 + z)^4)}
\end{eqnarray}

\begin{eqnarray}
\mid \bar{M}(^1S_0) \mid ^2=\frac{(-8\pi (8 + 3 \delta^2 - 8 z)
(\delta - z) (\delta + z))}{(9 s^2 (4 + \delta^2 - 4 z)^2 (-2 +
z)^2)}
\end{eqnarray}

With the chosen parameters mentioned above, the contributions to
the $\delta_1 c\bar c$ production cross section come from the
color-octet $^3P_0$, $^3P_1$ and $^3P_2$ are 0.18fb, 2.7fb and
0.87fb respectively. The contribution of the color-octet $^1S_0$
is negligible.

The total color-octet contribution to the cross section from
Fig.~\ref{fey2} is 8.3fb, and the sum of the color-octet
contributions from Fig.~\ref{feynman} and Fig.~\ref{fey2} is
9.9fb. This number could become substantially smaller if we use
the matrix element values given in \cite{omes1}.  In any case, the
color-octet contribution to the D-wave states production cross
sections should be of the same order as the color-singlet
contribution, because from the $\delta_J$ Fock state expansion in
eq.(\ref{expandfock}) and from Fig.~\ref{feynman} and
Fig.~\ref{fey2} it is easy to see that the color-octet and
color-singlet contributions are of the same order in both short
distance ($O(\alpha_s^2)$) and long distance ($O(v^4)$) parts.
Based on the above calculations including both the color-singlet
and color-octet contributions, we think
\begin{equation}
\label{dt}
\sigma(e^{+}+e^{-}\rightarrow\gamma^{*}\rightarrow\delta_{1}+c\bar{c})\approx
10~ {\rm fb}
\end{equation}
should be a reasonable estimate in NRQCD for the $\psi(3770)$
production rate when the S-D mixing is neglected.

\section{S-D mixing and $\psi(3770)$ production in  $e^+e^-$ annihilation and $B$ decay}

The calculated $\psi(3770)c\bar c$ production rate could be
significantly enhanced by the S-D mixing if the mixing angle is
large. With the calculated $J/\psi$ production cross section
$\sigma(e^{+}+e^{-}\rightarrow\gamma^{*}\rightarrow
J/\psi+c\bar{c})=148~fb$ \cite{liu1}, the $\psi(2S)$ production
cross section can be approximately obtained by the scale factor
$\mid R_{2S}(0) \mid^2/\mid R_{1S}(0) \mid^2$, and then we have
$\sigma(e^{+}+e^{-}\rightarrow\gamma^{*}\rightarrow
\psi(2S)+c\bar{c})=90~fb$ (the color-octet contributions to the
S-wave charmonia are small and negligible). If this estimate for
the $\psi(2S)$ makes sense, with the large S-D mixing angle
$\theta\approx \pm 40^o$ we would have
\begin{equation}
\label{dt3} \sigma(e^{+}+e^{-}\rightarrow
\psi(3770)+c\bar{c})\approx \sigma(e^{+}+e^{-}\rightarrow
\psi(2S)+c\bar{c})\times \tan^2\theta\approx 66~ {\rm fb}.
\end{equation}
This value is much larger than 10~fb, the value in eq.(\ref{dt})
obtained without S-D mixing in NRQCD.

However, as we already mentioned, the Belle observed double charm
production cross section for the $J/\psi$ has a much higher
value\cite{exdou}
\begin{equation}
\label{bpsi} \sigma(e^{+}+e^{-}\rightarrow
J/\psi+c\bar{c})=(0.87\pm 0.15\pm 0.12)~{\rm pb},
\end{equation}
which is larger than the theoretical expectation in NRQCD by a
factor of 6 (as a rough estimate we neglect the feed down
contribution from the $\psi(2S)$ and $\chi_{cJ}$ states in the
$J/\psi$ production cross section). If this also happens for the
$\psi(2S)$, which is very likely, we would expect
\begin{equation}
\label{bpsi'} \sigma(e^{+}+e^{-}\rightarrow
\psi(2S)+c\bar{c})\approx 530~{\rm fb}
\end{equation}
to be the observed double charm production cross section for the
$\psi(2S)$. Then with the large S-D mixing angle $\theta\approx
\pm 40^o$ we would have
\begin{equation}
\label{dt4} \sigma(e^{+}+e^{-}\rightarrow
\psi(3770)+c\bar{c})\approx \sigma(e^{+}+e^{-}\rightarrow
\psi(2S)+c\bar{c})\times \tan^2\theta\approx 387~ {\rm fb}.
\end{equation}
This is more than an order of magnitude larger than the value in
eq.(\ref{dt}) obtained without S-D mixing in NRQCD.

In any case, we see that the large S-D mixing angle $\theta\approx
\pm 40^o$ would result in a much higher cross section for the
$\psi(3770)$ double charm production than that without S-D mixing
or with small S-D mixing angle like $\theta\approx -10^o$. This is
due to the fact that the double charm production rate for a pure
$2^3S_1$ state is much higher than that for a pure $1^3D_1$ state
in NRQCD. We therefore suggest measuring the production cross
sections of $\psi(3770)+c\bar c$ and $\psi(3686)+c\bar c$ in the
continuum $e^+e^-$ annihilation at $\sqrt{s}=10.6$ GeV by Belle
and BaBar.

In contrast, in the $B$ inclusive decay to $\psi(3770)$, the decay
rate is expected to be insensitive to the S-D mixing in NRQCD,
because both the $2^3S_1$ and $1^3D_1$ final states produced in
$B$ decay are dominated by the color-octet intermediate states
with the same quantum numbers in the two cases and thus they may
have comparable production rates. The color-octet dominance in $B$
decay relies on two observations. The first is that in the
effective weak interactions the squared short distance coefficient
at the $b$ quark mass scale for the color-octet part is larger
than that for the color-singlet part by more than an order of
magnitude. The second is that the color-singlet S-wave
contribution is further suppressed by QCD radiative corrections
and therefore negligible (see \cite{bbyl}\cite{be}\cite{bmr} for
detailed discussions). Therefore, despite of certain uncertainties
related to the values of color-octet matrix elements and other
parameters, the qualitative features for the D-wave charmonium
production in $e^+e^-$ annihilation and $B$ decay should hold and
be tested.

\section{S-D mixing and $\psi(3770)$ E1 transitions}

As mentioned already, the small S-D mixing angle like
$\theta\approx -10^o$ is favored by the observed E1 transition
rates of $\psi(3686)$. However, as another independent check for
the mixing angle, measurements on the $\psi(3770)$ E1 transitions
will be also very useful. In ref.\cite{ding} the E1 transition
rates of $\psi(3770)$ were calculated for $\theta = 0^o, -10^o,
+30^o$, and the S-D interference effects were found to be
significant. Based on the same potential model we now estimate the
E1 transition widths for $\theta = \pm 40.4^o $ and find
\begin{equation}
\label{w1} \Gamma (\psi(3770)\rightarrow
\gamma\chi_{cJ})=386,~0.32,~66~KeV
\end{equation}
for J=0,1,2 with $\theta =-40.4^o $;~ and
\begin{equation}
\label{w2} \Gamma (\psi(3770)\rightarrow
\gamma\chi_{cJ})=52,~203,~28~KeV
\end{equation}
for J=0,1,2 with $\theta =+40.4^o $.

We see that the S-D interference effects in the $\psi(3770)$ E1
transitions are essential. In particular, the ratio
\begin{equation}
\label{r} R_{2/1}=\frac{\Gamma(\psi(3770)\rightarrow
\gamma\chi_{c2})}{\Gamma(\psi(3770)\rightarrow \gamma\chi_{c1})}
\end{equation}
will dramatically change from $\sim$ 0.04 (for $\theta\approx
0^o$) to $\sim$ 200 (for $\theta\approx -40^o$) due to the large
S-D interference effect. Similar discussions for $\theta\approx
-12^o$ and $+27^o$ can also be found in \cite{rosner}.

We hope these measurements can be performed at BES and CLEO-c in
the near future. They will be very helpful in clarifying the S-D
mixing problem for $\psi(3770)$ and $\psi(3686)$.

\section{Discussions and summary}

We first discuss about the uncertainties associated with the
color-octet matrix elements with different choices from
eqs.(13)-(15). According to the NRQCD velocity scaling rules, we
may have
\begin{equation}
\langle {\cal O}_1^{\delta_J}(^3D_J) \rangle \sim m_c^7v^7,~
\langle {\cal O}_8^{\delta_J}(^3P_1) \rangle \sim m_c^5v^7,~
\langle {\cal O}_8^{\delta_J}(^3S_1) \rangle \sim m_c^3v^7.
\end{equation}
If we use $\langle {\cal O}_1^{\delta_1}(^3D_1) \rangle$ as the
input parameter, $\langle {\cal O}_1^{\delta_1}(^3D_1)
\rangle=\frac{45N_c}{4\pi} \mid R_D''(0) \mid^2=0.16GeV^7$, we
would have the following matrix elements
\begin{equation}
\frac{\langle {\cal O}_1^{\delta_1}(^3D_1) \rangle}{m_c^7}\approx
\frac{\langle {\cal O}_8^{\delta_1}(^3P_1) \rangle}{m_c^5}\approx
\frac{\langle {\cal O}_8^{\delta_1}(^3S_1) \rangle}{m_c^3}=0.0094.
\end{equation}
Then the color-octet contributions to the cross section
$\sigma(e^+e^-\rightarrow(^{2S+1}L_J^{(8)})+c\bar{c}\rightarrow
\delta_1 + c\bar{c} )$ can be estimated to be
0.36~fb,~1.1~fb,~0.28~fb,~0.12~fb for
$^3S_1^{(8)},^3P_0^{(8)},^3P_1^{(8)},^3P_2^{(8)}$ color-octet
intermediate states respectively from Fig.~1; and
13.5~fb,~0.17~fb,~2.6~fb,~0.83~fb respectively from Fig.~2. The
total contribution to the cross section $\sigma(e^+e^-\rightarrow
\gamma^*\rightarrow \delta_1 + c\bar{c} )$ is about ~20~fb, a
factor of 2 larger than 10~fb, the value given in eq.(25) obtained
by using eqs.(13)-(15). The above results may demonstrate the
possible uncertainties associated with the color-octet matrix
elements, which are estimated according to the NRQCD velocity
scaling rules but with different choices for the input parameters.
We expect that these uncertainties do not change our main analysis
that the $\psi(3770)$ production in  $e^+e^-$ annihilation
receives much smaller contributions from the color-octet channels
than from the large S-D mixing.

We suggest performing the measurements on the $\psi(2S)$ and
$\psi(3770)$ production in the continuum $e^+e^-$ annihilation by
Belle and BaBar in order to check (i) if the $\psi(2S)+c\bar c$
cross section is as large as about 0.5pb, which would confirm the
S-wave (not only for 1S but also for 2S) charmonium production
enhancement via double $c\bar c$; (ii) if the cross section of
$\psi(3770)+c\bar c$ is comparable to that of $\psi(2S)+c\bar c$,
which would confirm the large S-D mixing angle; (iii) if the cross
section of $\psi(3770)+c\bar c$ is as small as say 10~fb, which
would favor the prediction in NRQCD. It is very likely that Belle
and BaBar will find a strong signal for $\psi(2S)+c\bar c$,
because experimentally Belle has found a large cross section for
the inclusive $\psi(2S)+X$ cross section, which is comparable to
that for $J/\psi+X$ \cite{belle}.  If a large $\psi(2S)+c\bar c$
cross section turned to be the case, then one can easily
distinguish between the large S-D mixing and the NRQCD prediction
(together with the small S-D mixing) by simultaneously measuring
the $\psi(3770)+c\bar c$ cross section.

In conclusion, we notice that the large decay rate observed by
Belle for $B^+\rightarrow\psi(3770)K^+$, which is comparable to
$B^+\rightarrow\psi(3686)K^+$, might indicate either an
unexpectedly large S-D mixing angle $|\theta|\approx 40^o$ or the
leading role of the color-octet mechanism in D-wave charmonium
production in $B$ decay. By calculating the production rate of
$\psi(3770)$ in the continuum $e^+e^-$ annihilation at
$\sqrt{s}=10.6$ GeV with these two possible approaches (i.e. the
large S-D mixing and the color-octet mechanism), we show that the
measurement for this process at Belle and BaBar may provide a
clear cut clarification for the two approaches. In addition, the
radiative E1 transition ratio $\Gamma(\psi(3770)\rightarrow
\gamma\chi_{c2})/\Gamma(\psi(3770)\rightarrow \gamma\chi_{c1})$
may dramatically change from $\sim$ 0.04 (for $\theta\approx 0^o$)
to $\sim$ 200 (for $\theta\approx -40^o$) due to the large S-D
interference effect, thus the E1 transition measurement of
$\psi(3770)$ at BES and CLEO-c will also be very useful in
clarifying this issue.

\section*{Acknowledgments}

We thank Z.G.He and Y.J.Zhang for help in checking some of the
calculations.  This work was supported in part by the National
Natural Science Foundation of China, the Education Ministry of
China, and the Beijing Electron Positron Collider National
Laboratory.

\section*{Appendix}

In this appendix, for $\delta_1$ and $\delta_2$ we give the
expressions of $S(z)$ and $\alpha(z)$ which are defined in
Eq.~(\ref{cross}).
\begin{eqnarray}
S_{\delta_1}&=&\frac{16\pi}{225 \delta^6s^4(\delta - z)(-2 +
z)^{10}z^7(\delta + z)}\{-4z\sqrt{\frac{(1 - z)(z^2-\delta^2)}{4
+ \delta^2 - 4z}}\nonumber\\
&\times&[-5529600\delta^{12} - 1382400\delta^{14} +
(42393600\delta^{10} + 38246400\delta^{12} +
5529600\delta^{14})z\nonumber\\
&-&(124600320\delta^8 + 260259840\delta^{10} +
109255680\delta^{12} + 10114560\delta^{14})z^2 \nonumber\\
&+&(145981440\delta^6 + 702873600\delta^8 + 687037440\delta^{10} +
175499520\delta^{12} \nonumber\\
&+& 10955520\delta^{14})z^3 - (106168320\delta^4 +
732979200\delta^6 + 1706803200\delta^8 \nonumber\\
&+& 1037391360\delta^{10} + 178859520\delta^{12} +
7741440\delta^{14})z^4 + (483655680\delta^4 \nonumber\\
&+& 1502760960\delta^6 + 2324674560\delta^8 + 992090880\delta^{10}
+ 121618560\delta^{12} \nonumber\\
&+& 3672000\delta^{14})z^5 -(106168320\delta^2 + 759152640\delta^4
+ 1448421376\delta^6 \nonumber\\
&+& 1866987008\delta^8 + 606618880\delta^{10} +
54528192\delta^{12} + 1110144\delta^{14})z^6 \nonumber\\
&+&(524943360\delta^2 + 98590720\delta^4 + 329316352\delta^6 +
803530752\delta^8 + 221747328\delta^{10} \nonumber\\
&+& 15245760\delta^{12} + 188688\delta^{14})z^7 + (70778880 -
1436221440\delta^2 + 1035042816\delta^4 \nonumber\\
&+& 448612352\delta^6 - 146389760\delta^8 - 42081920\delta^{10} -
2067024\delta^{12} + 26616\delta^{14})z^8 \nonumber\\
&-& (566231040 - 2742845440\delta^2 + 1070004224\delta^4 +
82320896\delta^6 - 40385536\delta^8 \nonumber\\
&-& 73280\delta^{10} + 262944\delta^{12} + 3180\delta^{14})z^9 +
(2235432960 - 3145580544\delta^2 \nonumber\\
&+& 493045760\delta^4 - 287114752\delta^6 - 34498880\delta^8 +
1811280\delta^{10} - 41160\delta^{12} \nonumber\\
&+& 630\delta^{14})z^{10} - (5315624960 - 1154056192\delta^2 +
773548544\delta^4 - 101372288\delta^6 \nonumber\\
&-& 12039424\delta^8 - 258940\delta^{10} + 5445\delta^{12}) z^{11}
+ (8075444224 + 1656672256\delta^2 \nonumber\\
&+& 1123221504\delta^4 + 50333664\delta^6 - 1196496\delta^8 +
110862\delta^{10} - 315\delta^{12})z^{12} \nonumber\\
&-& (8028487680 + 2410575872\delta^2 + 723781760\delta^4 +
38182256\delta^6 + 791960\delta^8 \nonumber\\
&-& 795\delta^{10}) z^{13} + (5278310400 + 1389795328\delta^2 +
236564032\delta^4 + 8583048\delta^6 \nonumber\\
&+& 16302\delta^8)z^{14} - (2310758400 + 445429248\delta^2 +
39874240\delta^4 + 590984\delta^6)z^{15} \nonumber\\
&+& (673812480 + 84606976\delta^2 + 3518976\delta^4)z^{16} -
(124616704 + 9082368\delta^2)z^{17} \nonumber\\
&+& 11939840z^{18}] + 15\delta^2(z-2)^4{\rm ln}[\frac{z\sqrt{4 +
\delta^2 - 4z} + 2\sqrt{(1 - z)(z^2-\delta^2)}}{z\sqrt{4 +
\delta^2 - 4z} - 2\sqrt{(1 - z)(z^2-\delta^2)}}]\nonumber\\
&\times&[23040\delta^{12} - (176640\delta^{10} +
69120\delta^{12})z + (519168\delta^8 + 585984\delta^{10} \nonumber\\
&+& 89856\delta^12)z^2 - (608256\delta^6 + 1620480\delta^8 +
799296\delta^{10} + 63936\delta^{12})z^3 \nonumber\\
&+& (442368\delta^4 + 1380864\delta^6 + 1912960\delta^8 +
578944\delta^{10} + 26880\delta^{12})z^4 \nonumber\\
&-&(724992\delta^4 + 378112\delta^6 + 924480\delta^8 +
227248\delta^{10} + 6144\delta^{12})z^5 \nonumber\\
&-& (1032192\delta^2 + 256000\delta^4 + 1437696\delta^6 +
22176\delta^8 - 46976\delta^{10} - 1056\delta^{12})z^6 \nonumber\\
&+& (2310144\delta^2 + 503808\delta^4 + 1262528\delta^6 +
108816\delta^8 - 11916\delta^{10} - 252\delta^{12})z^7 \nonumber\\
&-&(229376\delta^2 - 856576\delta^4 - 194944\delta^6 -
29976\delta^8 - 3592\delta^{10} - 42\delta^{12})z^8 \nonumber\\
&+& (294912 - 3428352\delta^2 - 1397248\delta^4 - 571184\delta^6 -
25516\delta^8 - 447\delta^{10})z^9 \nonumber\\
&-&(1392640 - 4259840\delta^2 - 583680\delta^4 - 215072\delta^6 -
3078\delta^8 + 21\delta^{10})z^{10} \nonumber\\
&+& (1998848 - 2720768\delta^2 + 20736\delta^4 - 16052\delta^6 +
345\delta^8)z^{11} \nonumber\\
&-& (1163264 - 1031936\delta^2 + 98816\delta^4 +
3554\delta^6)z^{12} + (284672 - 172800\delta^2 \nonumber\\
&+&22288\delta^4)z^{13} - (35840 - 5376\delta^2)z^{14} +
4096z^{15}]\}.
\end{eqnarray}

\begin{eqnarray}
\alpha_{\delta_1}&=&\frac{16\pi}{225 \delta^6s^4(\delta - z)(-2 +
z)^{10}z^7(\delta + z)}\{-4z\sqrt{\frac{(1 - z)(z^2-\delta^2)}{4
+ \delta^2 - 4z}}\nonumber\\
&\times& [5529600 \delta^{12} + 1382400 \delta^{14} - (46080000
\delta^{10} + 39168000 \delta^{12} + 5529600 \delta^{14}) z \nonumber\\
&+& (124600320 \delta^8 + 248832000 \delta^{10} + 105477120
\delta^{12} + 10114560 \delta^{14}) z^2 \nonumber\\
&-& (48660480 \delta^6 + 613171200 \delta^8 + 574632960
\delta^{10} + 155262720 \delta^{12} \nonumber\\
&+& 10955520 \delta^{14}) z^3 + (35389440 \delta^4 + 78028800
\delta^6 + 1309163520 \delta^8 \nonumber\\
&+& 755489280 \delta^{10} + 140951040 \delta^{12} + 7741440
\delta^{14}) z^4 - (117964800 \delta^4 \nonumber\\
&-& 359485440 \delta^6 + 1567180800 \delta^8 + 620524800
\delta^{10} + 82139520 \delta^{12} \nonumber\\
&+& 3672000 \delta^{14}) z^5 + (318504960 \delta^2 + 290734080
\delta^4 - 1587570688 \delta^6 \nonumber\\
&+& 1071024640 \delta^8 + 324301568 \delta^{10} + 29522496
\delta^{12} + 1110144 \delta^{14}) z^6 \nonumber\\
&-& (1032192000 \delta^2 + 235642880 \delta^4 - 3195049984
\delta^6 + 256432128 \delta^8 \nonumber\\
&+& 97444224 \delta^{10} + 4799424 \delta^{12} + 188688
\delta^{14}) z^7 - (-70778880 + 276234240 \delta^2 \nonumber\\
&+& 1370357760 \delta^4 + 4314056704 \delta^6 + 282550016 \delta^8
- 6669952 \delta^{10} + 647376 \delta^{12} \nonumber\\
&+& 26616 \delta^{14}) z^8 + (-566231040 + 6529515520 \delta^2 +
4711155712 \delta^4 \nonumber\\
&+& 4188711424 \delta^6 + 355571200 \delta^8 + 6214976 \delta^{10}
+ 832320 \delta^{12} + 3180 \delta^{14}) z^9 \nonumber\\
&-& (-2235432960 + 14312718336 \delta^2 + 6829299712 \delta^4 +
2844373504 \delta^6 \nonumber\\
&+& 199898048 \delta^8 + 4164432 \delta^{10} + 42696 \delta^{12} +
630 \delta^{14}) z^{10} + (-5315624960 \nonumber\\
&+& 15998746624 \delta^2 + 5588809216 \delta^4 + 1329557888
\delta^6 + 74936896 \delta^8 \nonumber\\
&+& 265900 \delta^{10} + 19515 \delta^{12}) z^{11} - (-8075444224
+ 10653200384 \delta^2 \nonumber\\
&+& 2855909376 \delta^4 + 440017440 \delta^6 + 16253328 \delta^8 +
284850 \delta^{10} + 315 \delta^{12}) z^{12} \nonumber\\
&+& (-8028487680 + 4372645888 \delta^2 + 953109376 \delta^4 +
93564688 \delta^6 \nonumber\\
&+& 3276232 \delta^8 + 795 \delta^{10}) z^{13} + (5278310400 -
1039542272 \delta^2 - 183272384 \delta^4 \nonumber\\
&-& 12182904 \delta^6 + 16302 \delta^8) z^{14} - (2310758400 -
70162944 \delta^2 - 11162432 \delta^4 \nonumber\\
&+& 590984 \delta^6) z^{15} + (673812480 + 33531904 \delta^2 +
3518976 \delta^4) z^{16} \nonumber\\
&-& (124616704 + 9082368 \delta^2) z^{17} + 11939840 z^{18}] \nonumber\\
&-& 15 \delta^2 (-2 + z)^4{\rm ln}[\frac{z\sqrt{4 + \delta^2 - 4z}
+ 2\sqrt{(1 - z)(z^2-\delta^2)}}{z\sqrt{4 +
\delta^2 - 4z} - 2\sqrt{(1 - z)(z^2-\delta^2)}}] \nonumber\\
&\times&[23040 \delta^{12} - (192000 \delta^{10} + 69120
\delta^{12}) z + (519168 \delta^8 + 507648 \delta^{10} \nonumber\\
&+& 89856 \delta^{12}) z^2 - (202752 \delta^6 + 1337856 \delta^8 +
533952 \delta^{10} + 63936 \delta^{12}) z^3 \nonumber\\
&+& (147456 \delta^4 - 463360 \delta^6 + 1415552 \delta^8 + 245120
\delta^{10} + 26880 \delta^{12}) z^4 \nonumber\\
&-& (61440 \delta^4 - 2630912 \delta^6 + 731072 \delta^8 + 13136
\delta^{10} + 6144 \delta^{12}) z^5 \nonumber\\
&-& (737280 \delta^2 + 219136 \delta^4 + 4452352 \delta^6 - 169760
\delta^8 + 38144 \delta^{10} - 1056 \delta^{12}) z^6 \nonumber\\
&+& (835584 \delta^2 - 724992 \delta^4 + 3959872 \delta^6 - 2256
\delta^8 + 10380 \delta^{10} - 252 \delta^{12}) z^7 \nonumber\\
&-& (1179648 - 3260416 \delta^2 - 4187648 \delta^4 + 1798528
\delta^6 - 46024 \delta^8 - 488 \delta^{10} \nonumber\\
&-& 42 \delta^{12}) z^8 + (3244032 - 10629120 \delta^2 - 7306752
\delta^4 + 69936 \delta^6 - 32804 \delta^8 \nonumber\\
&-& 465 \delta^{10}) z^9 - (3031040 - 13373440 \delta^2 - 6077440
\delta^4 - 218592 \delta^6 - 7418 \delta^8 \nonumber\\
&-& 21 \delta^{10}) z^{10} + (950272 - 8346624 \delta^2 - 2431744
\delta^4 - 61164 \delta^6 - 345 \delta^8) z^{11} \nonumber\\
&+& (57344 + 2462976 \delta^2 + 428416 \delta^4 + 3554 \delta^6)
z^{12} - (63488 + 259328 \delta^2 \nonumber\\
&+& 22288 \delta^4) z^{13} + (17408 - 5376 \delta^2) z^{14} - 4096
z^{15}]\}.
\end{eqnarray}

\begin{eqnarray}
{S}_{\delta_2}&=&\frac{128\pi}{45 \delta^6 s^4 (\delta - z) (-2 +
z)^{10} z^7 (\delta + z)}\{\sqrt{\frac{z^2-\delta^2}{(4 +
\delta^2 - 4 z) (1 - z)}}\nonumber\\
&\times&[1843200 \delta^{10} (4 + \delta^2)^2 z - 614400 \delta^8
(4 +
\delta^2) (72 + 101 \delta^2 + 15 \delta^4) z^2 \nonumber\\
&+& 30720 \delta^6 (10368 + 42816 \delta^2 + 38368 \delta^4 + 9574
\delta^6 + 679 \delta^8) z^3 \nonumber\\
&-& 5120 \delta^6 (390528 + 828736 \delta^2 + 519596 \delta^4 +
100453 \delta^6 + 5487 \delta^8) z^4 \nonumber\\
&+& 5120 \delta^4 (-110592 + 1058144 \delta^2 + 1531104 \delta^4 +
744394 \delta^6 + 115197 \delta^8 + 4869 \delta^{10}) z^5
\nonumber\\
&-& 1280 \delta^4 (-2955264 + 6395328 \delta^2 + 7124384 \delta^4
+ 2865924 \delta^6 + 364442 \delta^8 \nonumber\\
&+& 11889 \delta^{10}) z^6 + 256 \delta^2 (1105920 - 43674880
\delta^2 + 28767008 \delta^4 + 26401544 \delta^6 \nonumber\\
&+& 9368320 \delta^8 + 1006841 \delta^{10} + 24907 \delta^{12})
z^7 - 64 \delta^2 (28139520 - 296599040 \delta^2 \nonumber\\
&+& 61689856 \delta^4 + 44813600 \delta^6 + 15834776 \delta^8 +
1516544 \delta^{10} + 27059 \delta^{12}) z^8 \nonumber\\
&+& 32 (-2949120 + 160276480 \delta^2 - 617186304 \delta^4 +
56704000 \delta^6 + 10167648 \delta^8 \nonumber\\
&+& 7112432 \delta^{10} + 718126 \delta^{12} + 6753 \delta^{14})
z^9 + 16 (47185920 - 527851520 \delta^2 \nonumber\\
&+& 729945600 \delta^4 - 149188736 \delta^6 + 13544832 \delta^8 +
47360 \delta^{10} - 144180 \delta^{12} + 2483 \delta^{14})
z^{10}\nonumber\\
&-& 8 (365363200 - 1046585344 \delta^2 + 224047104 \delta^4 -
415587456 \delta^6 - 2576256 \delta^8 \nonumber\\
&+& 1710200 \delta^{10} + 27764 \delta^{12} + 635 \delta^{14})
z^{11} + 4 (1772748800 - 990953472 \delta^2\nonumber\\
&-&  665257472 \delta^4 - 664114560 \delta^6 - 29465408 \delta^8 +
277460 \delta^{10} - 21175 \delta^{12} + 210 \delta^{14})
z^{12}\nonumber\\
&-&  8 (1473757184 + 199317504 \delta^2 - 234019072 \delta^4 -
150230864 \delta^6 - 7587160 \delta^8 \nonumber\\
&-& 124303 \delta^{10} + 300 \delta^{12}) z^{13} - 4 (-3491954688
- 1095510016 \delta^2 + 45265792 \delta^4\nonumber\\
&+&  71393200 \delta^6 + 2794952 \delta^8 - 18599 \delta^{10} +
105 \delta^{12}) z^{14} - 4 (2995691520 \nonumber\\
&+& 946406400 \delta^2 + 85150144 \delta^4 - 3895992 \delta^6 +
95674 \delta^8 + 425 \delta^{10}) z^{15} \nonumber\\
&+& 8 (930959360 + 243791616 \delta^2 + 24343872 \delta^4 + 756392
\delta^6 + 4617 \delta^8) z^{16}\nonumber\\
&-&  96 (34273280 + 6615552 \delta^2 + 442968 \delta^4 + 5637
\delta^6) z^{17} \nonumber\\
&+& 1024 (956608 + 116158 \delta^2 + 3367 \delta^4) z^{18} - 16384
(10698 + 631 \delta^2) z^{19} + 14581760 z^{20}]\nonumber\\
&+&  [1843200 \delta^{12} (4 + \delta^2) - 614400 \delta^{10} (72
+ 89 \delta^2 + 15 \delta^4) z \nonumber\\
&+& 61440 \delta^8 (1296 + 4228 \delta^2 + 2819 \delta^4 + 342
\delta^6) z^2\nonumber\\
&-&  15360 \delta^8 (25440 + 42856 \delta^2 + 20593 \delta^4 +
1869 \delta^6) z^3\nonumber\\
&+&  7680 \delta^6 (-25344 + 104048 \delta^2 + 120884 \delta^4 +
48728 \delta^6 + 3391 \delta^8) z^4 \nonumber\\
&-& 3840 \delta^6 (-283392 + 218448 \delta^2 + 197300 \delta^4 +
78807 \delta^6 + 4272 \delta^8) z^5 \nonumber\\
&+& 3840 \delta^4 (46080 - 725888 \delta^2 + 107264 \delta^4 +
74942 \delta^6 + 44495 \delta^8 + 1901 \delta^{10}) z^6
\nonumber\\
&-& 1920 \delta^4 (540672 - 2251264 \delta^2 + 11080 \delta^4 -
37482 \delta^6 + 35347 \delta^8 + 1209 \delta^{10}) z^7
\nonumber\\
&+& 480 \delta^2 (-98304 + 5812224 \delta^2 - 9475072 \delta^4 +
31856 \delta^6 - 315944 \delta^8 + 40476 \delta^{10}\nonumber\\
&+&  1147 \delta^{12}) z^8 - 480 \delta^2 (-540672 + 9385984
\delta^2 - 7280768 \delta^4 + 362224 \delta^6 \nonumber\\
&&- 179332 \delta^8 + 9230 \delta^{10} + 229 \delta^{12}) z^9 +
240 \delta^2 (-2670592 + 20128768 \delta^2 - 8820608
\delta^4\nonumber\\
&+&  831568 \delta^6 - 110696 \delta^8 + 3933 \delta^{10} + 85
\delta^{12}) z^{10} - 60 \delta^2 (-15597568 + 59650048
\delta^2\nonumber\\
&-&  18438144 \delta^4 + 1779904 \delta^6 - 84800 \delta^8 + 2921
\delta^{10} + 49 \delta^{12}) z^{11} + 30 \delta^2 (-30146560
\nonumber\\
&+& 61106176 \delta^2 - 17407488 \delta^4 + 980688 \delta^6 -
26636 \delta^8 + 640 \delta^{10} + 7 \delta^{12})
z^{12}\nonumber\\
&-&  15 \delta^2 (-41222144 + 41398272 \delta^2 - 13922048
\delta^4 + 210896 \delta^6 - 9404 \delta^8 + 19 \delta^{10})
z^{13}\nonumber\\
&-&  15 \delta^2 (20840448 - 7954432 \delta^2 + 4165888 \delta^4 +
19040 \delta^6 + 1242 \delta^8 + 7 \delta^{10}) z^{14}\nonumber\\
&+&  15 \delta^2 (7962624 - 230400 \delta^2 + 829440 \delta^4 +
6508 \delta^6 + 69 \delta^8) z^{15} \nonumber\\
&-& 30 \delta^2 (1140736 + 139136 \delta^2 + 48432 \delta^4 + 219
\delta^6) z^{16} + 240 \delta^2 (29312 + 4064 \delta^2
\nonumber\\
&+& 321 \delta^4) z^{17} - 1920 \delta^2 (488 + 41 \delta^2)
z^{18} + 61440 \delta^2 z^{19}]\nonumber\\
&\times&\ln\frac{z\sqrt{4 + \delta^2 - 4 z} + 2 \sqrt{(1 -
z)(-\delta^2 + z^2)}} {z\sqrt{4 + \delta^2 - 4 z} - 2 \sqrt{(1 -
z)(-\delta^2 + z^2)}}\}.
\end{eqnarray}

\begin{eqnarray}
\alpha_{\delta_2}&=&\frac{128\pi}{45 \delta^6 s^4 (\delta - z) (-2
+ z)^{10} z^7 (\delta + z)}\{\sqrt{\frac{z^2-\delta^2}{(4 +
\delta^2 - 4 z) (1 - z)}}\nonumber\\
&\times&[-614400 \delta^{10} (4 + \delta^2) (4 + 3 \delta^2) z +
614400 \delta^8 (96 + 164 \delta^2 + 103 \delta^4 + 15 \delta^6)
z^2\nonumber\\
& -& 10240 \delta^6 (10368 + 47488 \delta^2 + 39440 \delta^4 +
17302 \delta^6 + 2037 \delta^8) z^3 \nonumber\\
&+& 5120 \delta^6 (124032 + 342400 \delta^2 + 174148 \delta^4 +
55779 \delta^6 + 5487 \delta^8) z^4 \nonumber\\
&-& 5120 \delta^4 (-55296 + 269984 \delta^2 + 717792 \delta^4 +
241766 \delta^6 + 57183 \delta^8 + 4869 \delta^{10}) z^5
\nonumber\\
&+& 1280 \delta^4 (-1480704 + 548416 \delta^2 + 3899488 \delta^4 +
895740 \delta^6 + 153438 \delta^8 + 11889 \delta^{10})
z^6\nonumber\\
&-&  256 \delta^2 (-368640 - 21308160 \delta^2 - 10322464 \delta^4
+ 18164600 \delta^6 + 2826944 \delta^8 \nonumber\\
&+& 319283 \delta^{10} + 24907 \delta^{12}) z^7 + 64 \delta^2
(-19292160 - 131187200 \delta^2 - 97682944 \delta^4 \nonumber\\
&+& 49654496 \delta^6 + 4871624 \delta^8 + 221640 \delta^{10} +
27059 \delta^{12}) z^8 \nonumber\\
&-& 32 (2949120 - 186982400 \delta^2 - 195251200 \delta^4 -
189128192 \delta^6 + 58393888 \delta^8 \nonumber\\
&+& 3168624 \delta^{10} - 155198 \delta^{12} + 6753 \delta^{14})
z^9 - 16 (-47185920 + 998072320 \delta^2 \nonumber\\
&-& 51746304 \delta^4 + 132896384 \delta^6 - 73729792 \delta^8 -
2617408 \delta^{10} + 269788 \delta^{12}\nonumber\\
& +& 2483 \delta^{14}) z^{10} + 8 (-365363200 + 3412402176
\delta^2 - 815620096 \delta^4 - 165636992 \delta^6 \nonumber\\
&-& 91975232 \delta^8 - 2604552 \delta^{10} + 185516 \delta^{12} +
635 \delta^{14}) z^{11} - 4 (-1772748800\nonumber\\
& +& 8014970880 \delta^2 - 1588918784 \delta^4 - 462642816
\delta^6 - 81292032 \delta^8 - 1714116 \delta^{10}\nonumber\\
& +& 24647 \delta^{12} + 210 \delta^{14}) z^{12} + 8 (-1473757184
+ 3323742208 \delta^2 - 346701568 \delta^4 \nonumber\\
&-& 104122672 \delta^6 - 10199336 \delta^8 - 237265 \delta^{10} +
3020 \delta^{12}) z^{13} - 4 (-3491954688 \nonumber\\
&+& 3865319424 \delta^2 - 63264896 \delta^4 - 36776144 \delta^6 -
2587160 \delta^8 + 24425 \delta^{10}\nonumber\\
& +& 105 \delta^{12}) z^{14} - 4 (2995691520 - 1482234880 \delta^2
- 56928320 \delta^4 - 12344 \delta^6 \nonumber\\
&-& 244390 \delta^8 + 425 \delta^{10}) z^{15} + 8 (930959360 -
150302976 \delta^2 - 7609536 \delta^4 \nonumber\\
&-& 529880 \delta^6 + 4617 \delta^8) z^{16} - 96 (34273280 +
217600 \delta^2 + 32792 \delta^4 + 5637 \delta^6) z^{17}
\nonumber\\
&+& 1024 (956608 + 61294 \delta^2 + 3367 \delta^4) z^{18} - 16384
(10698 + 631 \delta^2) z^{19} \nonumber\\
&+& 14581760 z^{20}] - [614400 \delta^{12} (4 + 3 \delta^2) -
614400 \delta^{10} (24 + 31 \delta^2 + 15 \delta^4) z \nonumber\\
&+& 61440 \delta^8 (432 + 1604 \delta^2 + 909 \delta^4 + 342
\delta^6) z^2 - 15360 \delta^8 (7968 + 19112 \delta^2 \nonumber\\
&+& 5535 \delta^4 + 1869 \delta^6) z^3 + 7680 \delta^6 (-11520 +
21648 \delta^2 + 68348 \delta^4 + 9120 \delta^6 \nonumber\\
&+& 3391 \delta^8) z^4 - 3840 \delta^6 (-128256 - 39504 \delta^2 +
165980 \delta^4 + 5417 \delta^6 + 4272 \delta^8) z^5 \nonumber\\
&+& 3840 \delta^4 (39936 - 299136 \delta^2 - 227520 \delta^4 +
146914 \delta^6 - 4783 \delta^8 + 1901 \delta^{10}) z^6\nonumber\\
& -& 1920 \delta^4 (491520 - 740864 \delta^2 - 794120 \delta^4 +
198970 \delta^6 - 13267 \delta^8 + 1209 \delta^{10}) z^7
\nonumber\\
&+& 480 \delta^2 (-98304 + 5386240 \delta^2 - 1953792 \delta^4 -
3425904 \delta^6 + 424008 \delta^8 - 30108 \delta^{10}\nonumber\\
& +& 1147 \delta^{12}) z^8 - 480 \delta^2 (-442368 + 8620032
\delta^2 - 530304 \delta^4 - 2668784 \delta^6 + 175028
\delta^8\nonumber\\
& -& 9350 \delta^{10} + 229 \delta^{12}) z^9 + 240 \delta^2
(-1753088 + 17685504 \delta^2 - 52352 \delta^4 - 3219728
\delta^6\nonumber\\
& +& 102032 \delta^8 - 2797 \delta^{10} + 85 \delta^12) z^{10} -
60 \delta^2 (-7995392 + 48058368 \delta^2 - 1542144
\delta^4\nonumber\\
& -& 6071104 \delta^6 + 65680 \delta^8 + 135 \delta^{10} + 49
\delta^{12}) z^{11} + 30 \delta^2 (-11141120 + 43489280
\delta^2\nonumber\\
& -& 4092416 \delta^4 - 4255696 \delta^6 + 700 \delta^8 + 632
\delta^{10} + 7 \delta^{12}) z^{12} - 15 \delta^2
(-8323072\nonumber\\
& +& 26431488 \delta^2 - 3950848 \delta^4 - 2012880 \delta^6 -
5492 \delta^8 + 173 \delta^{10}) z^{13} + 15 \delta^2
(196608\nonumber\\
& +& 5984256 \delta^2 - 673536 \delta^4 - 272864 \delta^6 - 182
\delta^8 + 7 \delta^{10}) z^{14} - 15 \delta^2 (2064384\nonumber\\
& +& 1408000 \delta^2 + 93184 \delta^4 - 13620 \delta^6 + 69
\delta^8) z^{15} + 30 \delta^2 (587776 + 184704
\delta^2\nonumber\\
& +& 24944 \delta^4 + 219 \delta^6) z^{16} - 240 \delta^2 (21632 +
4192 \delta^2 + 321 \delta^4) z^{17} \nonumber\\
&+& 1920 \delta^2 (440 + 41
\delta^2) z^{18} - 61440 \delta^2 z^{19}]\nonumber\\
&\times&\ln\frac{z\sqrt{4 + \delta^2 - 4 z} + 2 \sqrt{(1 -
z)(-\delta^2 + z^2)}} {z\sqrt{4 + \delta^2 - 4 z} - 2 \sqrt{(1 -
z)(-\delta^2 + z^2)}}\}.
\end{eqnarray}

\newpage

\newpage

\end{document}